\RequirePackage{fix-cm}

\documentclass[smallextended]{svjour3}       

\smartqed  
\usepackage{graphicx}
\usepackage[authoryear,sectionbib,sort]{natbib}

 \journalname{}
\begin{document}

\title{Complex ecological communities and the emergence of island species area relationships
}

\titlerunning{Complex ecological communities and island SARs}        

\author{Ankit Vikrant         \and
        Martin Nilsson Jacobi 
}


\institute{Ankit Vikrant \at
              Department of Space, Earth and Environment, Chalmers University of Technology, Maskingränd 2, 412 58 Gothenburg, Sweden \\
              \email{ankitvikrant74@gmail.com}           
           \and
           Martin Nilsson Jacobi \at
              Department of Space, Earth and Environment, Chalmers University of Technology, Maskingränd 2, 412 58 Gothenburg, Sweden
}

\date{Received: date / Accepted: date}

\maketitle

\begin{abstract}
It has been a century since the species-area relationship (SAR) was first proposed as a power law to explain how species richness scales with area. There have been many attempts to explain the origin of this predominant form. Apart from the power law, numerous empirical studies also report a semi-log form of the SAR, but very few have addressed its incidence. In this work, we test whether these relationships could emerge from the assembly of large random communities on island-like systems. We reformulate the generalized Lotka-Volterra model by introducing an area parameter that determines the species richness of the assembled communities. Our analysis demonstrates that the two most widely reported relationship forms can emerge due to differences in immigration rates and skewness towards weak interactions. We particularly highlight the incidence of the semi-log SAR for low immigration rates from a source pool, which is consistent with several previous empirical studies. The two SAR forms might show good fits to data over a large span of areas but a power-law overestimates species richness on smaller islands in remote archipelagoes.  

\end{abstract}

\section{Introduction}
\label{intro}
The species-area relationship (SAR) is arguably the most widely studied scaling law in ecology, having received empirical support from numerous studies spanning different geographical regions and taxa (\cite{lomolino2001towards, drakare2006imprint}). The predominant power law form of the SAR was first described by O. Arrhenius in 1921 (\cite{arrhenius1921species}). It related the number of species $S$ to the area of a habitat $A$ as $S \sim A^z$, where the exponent $z$ varies widely between 0 and 1 (\cite{drakare2006imprint}). A quantitative meta-analysis of a large number of SAR studies estimated its average value as 0.27 (\cite{drakare2006imprint}). The power law was contested by a semi-log relationship in 1922, that advocated the form $S \sim z \log(A)$ (\cite{gleason1922relation}). While the power law relationship is more widely reported, the semi-log SAR has also found support from numerous studies (\cite{lomolino2001towards, drakare2006imprint}). There have been attempts to explain the power law form based on species distributions (\cite{coleman1981random, leitner1997nested, picard2004species, vsizling2004power}), abundance distributions (\cite{preston1948commonness}) or population dynamics through constraints on immigration (\cite{durrett1996spatial, bastolla2001diversity}). The prevalence of SARs has also been attributed to the combined effects of widely observed abundance distributions and the fact that individuals from the same species cluster together (\cite{martin2006origin}). The semi-log relationship can be recovered from the power law SAR in some limit using species-incidence functions that depend on colonization and extinction rates (\cite{ovaskainen2003species}). However, there is no unified framework to explain the emergence of these competing SARs.

These scaling relationships are emergent in that they could be described by coarse-grained dynamics of large communities at the species level without reference to finer details and properties of individual organisms. Understanding the assembly of large communities could therefore underpin mechanisms that shape these scaling laws. The analysis of large systems has benefitted from many emerging approaches in the recent decades. In 1972, P.W. Anderson influenced the philosophy of science by suggesting that `more is different' (\cite{anderson1972more}), based on accumulating evidence from various disciplines. This means that the properties of a collective composed of many parts could be drastically different from the parts themselves. In the same year, R. May used random matrix theory to show that large ecosystems become unstable when their complexity increases beyond a threshold (\cite{may1972will}), which contradicted the prevailing notion that diversity increases stability. May's analytical results showed that one cannot have indefinite stability in large and complex ecosystems with many interactions. There is a limit beyond which an ecosystem is not resilient to small perturbations and can exhibit large fluctuations in the population abundances of the constituent species. He defined complexity in terms of connectance and interaction strength of the random matrix that encodes species interactions. 

The complex dynamics of such random interaction networks can be modelled using the Generalized Lotka-Volterra (GLV) equations. This model has been employed to uncover theoretical results ranging from identification of structural properties that affect coexistence (\cite{servan2018coexistence}) to the study of generic assembly patterns that are consistent across network structures (\cite{bunin2017ecological, barbier2018generic}). Some recent studies have investigated the distribution of number of coexisting species that results from GLV dynamics of much larger species pools (\cite{servan2018coexistence}). Others have even explored the progression and boundaries of extinction in large ecosystems (\cite{pettersson2020predicting}). These studies depart from identifying constraints on parameters that result in complete co-existence of all species. When the interaction strength is increased beyond the regime where all species co-exist, the system wades through a phase characterized by single-species extinctions (\cite{pettersson2020predicting}). May's stability limit marks the end of this phase beyond which no stable equilibria exist.

We hypothesize that a modified GLV model accounting for spatial scaling could exhibit SARs through the assembly of random communities of different sizes. Our analysis relies on introducing an area parameter to the GLV equations to test these questions. We explore a large part of this area parameter space to recover different number of surviving species beyond the regime of complete co-existence. By further allowing demographic immigration in the modified GLV model, we demonstrate that the two widely reported forms of the SAR stem from differences in immigration rates and the skewness towards weak interactions. We discuss the implications of our results in the context of island systems. The differences in the two SAR forms are more significant for smaller islands especially on distant archipelagoes, which we describe in \ref{sec_eg} and Supplementary appendix S2 using data from empirical studies (\cite{diamond1972biogeographic, diamond1976species, whittaker2014functional, gooriah2020species}). 

\section{Methods}

\subsection{Generalized Lotka-Volterra with spatial scaling}

\begin{figure}[!htb]
	\centering
	\includegraphics[width =\linewidth]{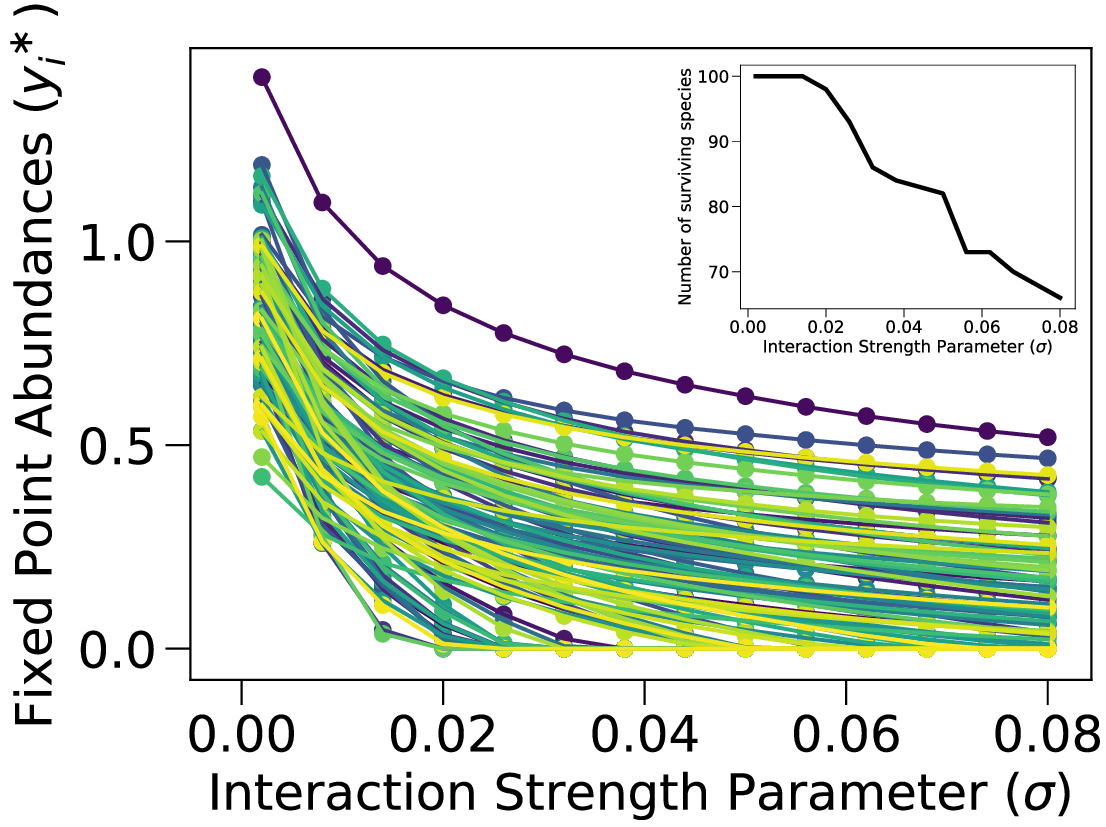}
	\caption{ Equilibrium abundances of a competitive community of 100 species for increasing values of the interaction strength parameter ($\sigma$). Each set of vertical dots represents an assembled community corresponding to a given value of $\sigma$. The bold black line in the inset traces the corresponding number of surviving species. Note that higher values of $\sigma$ correspond to more extinctions. The interaction strengths, growth rates and carrying capacities are chosen from normal distributions with means -1, +1 and +1 respectively. The standard deviation is set to 0.2 for each of these. }
\label{fig1}
\end{figure}

In its usual form, the GLV model describes the dynamics of species with densities $y_i$ through the following equations:\\
\begin{equation} \label{eq1}
	\frac{dy_i}{dt} = r_i y_i(1 - \frac{y_i}{K_i}) + \sigma y_i \sum_{j \neq i} B_{ij}y_j\\
\end{equation}
where $K_i$ and $r_i$ denote the carrying capacity and growth rate of $i^{th}$ species. $B_{ij}$ expresses pairwise inter-specific interaction strengths between species $i$ and any other species $j$. The full matrix B contains information about all possible pairwise interaction strengths between species. Equation \ref{eq1} implies that in the absence of interactions, each species grows to its carrying capacity $K_i$. $\sigma$ is the interaction strength parameter that scales all pairwise interaction strengths and consequently the variance of the interaction matrix $B$.

For a given value of $\sigma$ below May's limit, the system eventually relaxes to a stable equilibrium that represents an assembled community where species densities are resilient to small perturbations (Fig. 1). The equilibrium densities of species at this stable fixed point could either be zero or positive. Fixed points without any extinctions are called feasible solutions but these are of little concern to us since we are interested in communities with different number of surviving species assembled from a species pool.

To address a spatial scaling perspective, we first argue for a physical interpretation of the parameter $\sigma$. First note that $\sigma$ operates by scaling all the interaction strengths within the ecosystem by the same magnitude. This implies that an increase in this parameter reflects increased encounter rates between various species. We posit that $\sigma$ behaves like the inverse of area such that larger values this parameter correspond to smaller areas. 

More specifically, we replace densities in the GLV model by absolute abundances $x_i$ and area. Without the interaction strength, the equations are:
\begin{equation}\label{eq2}
	\frac{dx_i}{dt} = r_ix_i(1 - \frac{x_i}{K_i}) + \frac{x_i A_0}{A} \sum_{j \neq i} B_{ij}x_j\\
\end{equation}
where $A_0$ parameterises this model for a given ecosystem. We set this parameter equal to 1 from hereon. The carrying capacity $K_i$ is now an absolute quantity instead of a density, which explains why an area factor does not appear in that term. To account for the effect of decreasing areas on carrying capacities, we rewrite \ref{eq2} as:

\begin{equation}\label{eq2_1}
	\frac{dx_i}{dt} = r_ix_i(1 - \frac{x_i}{K_i (\frac{A}{A_{init}})^\gamma }) + \frac{x_i }{A} \sum_{j \neq i} B_{ij}x_j\\
\end{equation}

where $A_{init}$ is the area corresponding to the first extinction. We fix $\gamma = 0.25$ for the analysis described in this paper ( In general, $\gamma < 0.5$ is consistent with the results that we report) such that the carrying capacities are scaled weakly as compared to the scaling of interspecies interaction strengths. Equivalently, the self-interactions scale weakly with changes in area. This assumption is consistent with the fact that individuals of the same species cluster, which has also been widely echoed in various works explaining the power law SARs (\cite{plotkin2000species, martin2006origin}). The self-interactions already exist at higher levels, and therefore change less drastically as compared to interactions with other species.

This set of equations is similar to the usual GLV equations but now the interaction strength parameter $\sigma$ is replaced by the inverse of area. When working with abundances, it is more natural to argue that increase in the interaction strength parameter is analogous to decrease in area, which increases the encounter rates between species. 

We aim to describe a scenario where a regional pool of species is available to colonise different islands in a region (\cite{kessler2015generalized}). For an island defined by its area, the dynamics resulting from our model culminates in a final community where some species from the regional pool might not be feasible. Islands of different sizes yield communities with different compositions as a consequence. We use our model to simulate ecosystem dynamics as follows:\\
\begin{enumerate}
\item We pick entries of the interaction matrix $B_{ij}$ from a normal distribution that is symmetric around a negative mean (We fix mean = -1 and standard deviation = 0.2) . 
\item The growth rates $r_i$ are drawn from a normal distribution with mean = 1 and standard deviation = 0.2. The constraints on interactions and growth rates describe a community of competitive species.
\item The carrying capacities $K_i$ are normally distributed with mean = 500 and standard deviation = 30. The parameter $\gamma$ is fixed as 0.25. The choice of  $B_{ij}, r_i$ and $K_i$ allows for a large range of areas for which the system relaxes to stable equilibria. 
\item Starting from an initial area, the number of surviving species is plotted against successively smaller island areas $A$. \\
\end{enumerate}

We are primarily interested in investigating the properties and processes of community assembly that could possibly influence SARs using our spatially-implicit model. In all cases that we describe, we only show comparisons between the power-law and semi-log relationship forms. We perform a non-linear least-squares (NLSQ) analysis to fit and compare these forms using the least\_squares function in `scipy.optimize' package. This function implements the Trust Region Reflective algorithm described in \cite{branch1999subspace}. We also plot the linear regression of the corresponding better form for each of the cases. If there are considerable differences between the parameter estimates from the linear regression and the NLSQ analysis (this is the case only for the power-law estimates from an empirical dataset with few islands (\cite{whittaker2014functional}) ) , then we perform model averaging using the R package `sars' (\cite{matthews2019sars}) to discern the better fit. 

Analogous to the scenario of increasing $\sigma$, the system relaxes to a unique stable fixed point when the area parameter is above a certain threshold. We obtain the number of surviving species from the fixed point for each value of the area parameter (Fig. 1). We hypothesize that the different number of surviving species obtained by varying the area parameter result in widely reported SARs. These relationships are usually studied for one type of species or species that are placed in the same trophic level. This is congenial to our choice of a competitive interaction matrix. A competitive system could represent functional groups such as pollinators that compete for some common resources. A competitive GLV model with demographic noise has been shown to reproduce neutral island theories of Wilson-MacArthur and Hubbell (\cite{kessler2015generalized}). The power-law SAR has also been recovered from a spatially-explicit extension of the Lotka-Volterra competition model that allowed migration between patches (\cite{o2019metacommunity}).

\subsection{ Spatial scaling patterns with immigration}

Immigration slows down the decline in number of surviving species through introduction of new species (\cite{macarthur1963equilibrium}) or by delaying extinctions through incoming individuals of existing species (demographic immigration) (\cite{brown1977turnover}). What effects do different levels of immigration have on spatial scaling patterns in our model ecosystem? To address this, we redefine our GLV model with an additional term for demographic immigration:

\begin{equation}\label{eq3}
\frac{dx_i}{dt} = r_ix_i(1 - \frac{x_i}{K_i (\frac{A}{A_{init}})^\gamma}) + \frac{x_i}{A} \sum_{j \neq i} B_{ij}x_j + \lambda e^{-\beta / \sqrt{A} }  
\end{equation}

The last term represents the immigration rate. This term has a negligible contribution for smaller values of area, where a species may go extinct without support from the growth and interaction terms. A species is considered extinct in our simulations if its abundance falls below $10^{-5}$. As the area of an ecosystem shrinks, its distance from other patches also increases. This reduces the possibility of immigrants entering the ecosystem. The immigration term in the above equation has an exponential function that represents varying levels of demographic rescue (\cite{brown1977turnover}) as a function of area. For large values of area, $\lambda$ is the effective immigration rate. $\beta$ is a constant in the exponential function analogous to the characteristic length scale in the spatially extended GLV model described in \cite{o2019metacommunity}. We fixed $\beta = 1000$ and compared the results for different values of $\lambda$. 

We also consider interaction networks with more realistic connectances and distributions of interactions between species. Real food-webs are characterized by many weak and few strong interactions that also endow stability (\cite{mccann1998weak}). We study how the preponderance of weak interactions influences SARs. We use interactions drawn from exponential distributions that represent communities with varying skewness towards weak interactions. The rate parameter of the exponential distribution serves as a measure of this skew. 

In \ref{sec_eg}, we discuss our results in the light of two related empirical studies (\cite{diamond1972biogeographic, diamond1976species}) that exemplify the dependence of SARs on immigration rates. Both studies investigated bird diversity in the Southwest Pacific but differ in terms of their remoteness from the 'source island' of New Guinea. Our findings for low immigration rates (equivalently remote archipelagoes) are also consistent with the data from the Andaman and Azores Islands ( \cite{whittaker2014functional, gooriah2020species, gooriah2020data}, See Supplementary Appendix S2). 

The dataset used in (\cite{diamond1972biogeographic, diamond1976species}) has islands with areas spanning over six orders of magnitude, conclusively differentiating between competing forms of the SAR. These studies exclude `isolated' islands from their analysis, that are far from large islands within the archipelago. Speciation might influence the assembled communities especially on islands with fewer species. Islands whose avifaunas have not reached equilibrium are not included either. These are recolonized volcanic islands and islands that have undergone overall size contraction or modification of connecting land-bridges in the past c. 10,000 years. 

 \begin{figure}[!htb]
	\centering
	\includegraphics[width =\linewidth]{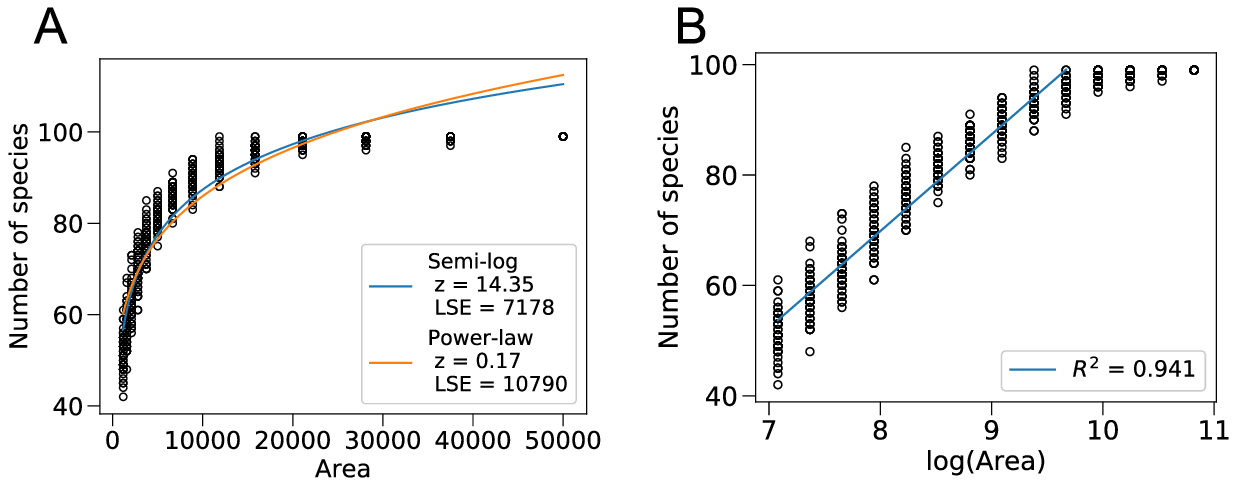}
	\caption{ Species-area plots generated through 50 realizations of interaction matrix with mean = -1 and variance = 0.2. $A_{init}$ = 50000. (A) The semi-log form shows a better fit. (B) The corresponding linear regression on a semi-log plot that shows an obvious upper asymptote.}
	\label{fig2}
\end{figure}

\section{Results}

Fig. 2 corresponds to the simplest case of an ecosystem with full connectance and no immigration. Starting with 100 species, we plot the number of surviving species for island areas where at least one species goes extinct. The SAR is best represented by a semi-log function through our model. The curve saturates at an upper asymptote for very high values of the area parameter (Fig. 2). 

The slope of the semi-log SAR varies with changes in the means of interactions and growth rates. It is also worth noting that for an intermediate range of areas, even the log-log plot could show a misleadingly good fit for a power law SAR (Fig. 2).

\subsection{ What determines a log-log or a semi-log SAR?}

 \begin{figure}[!htb]
    \centering
    	\includegraphics[width =\linewidth]{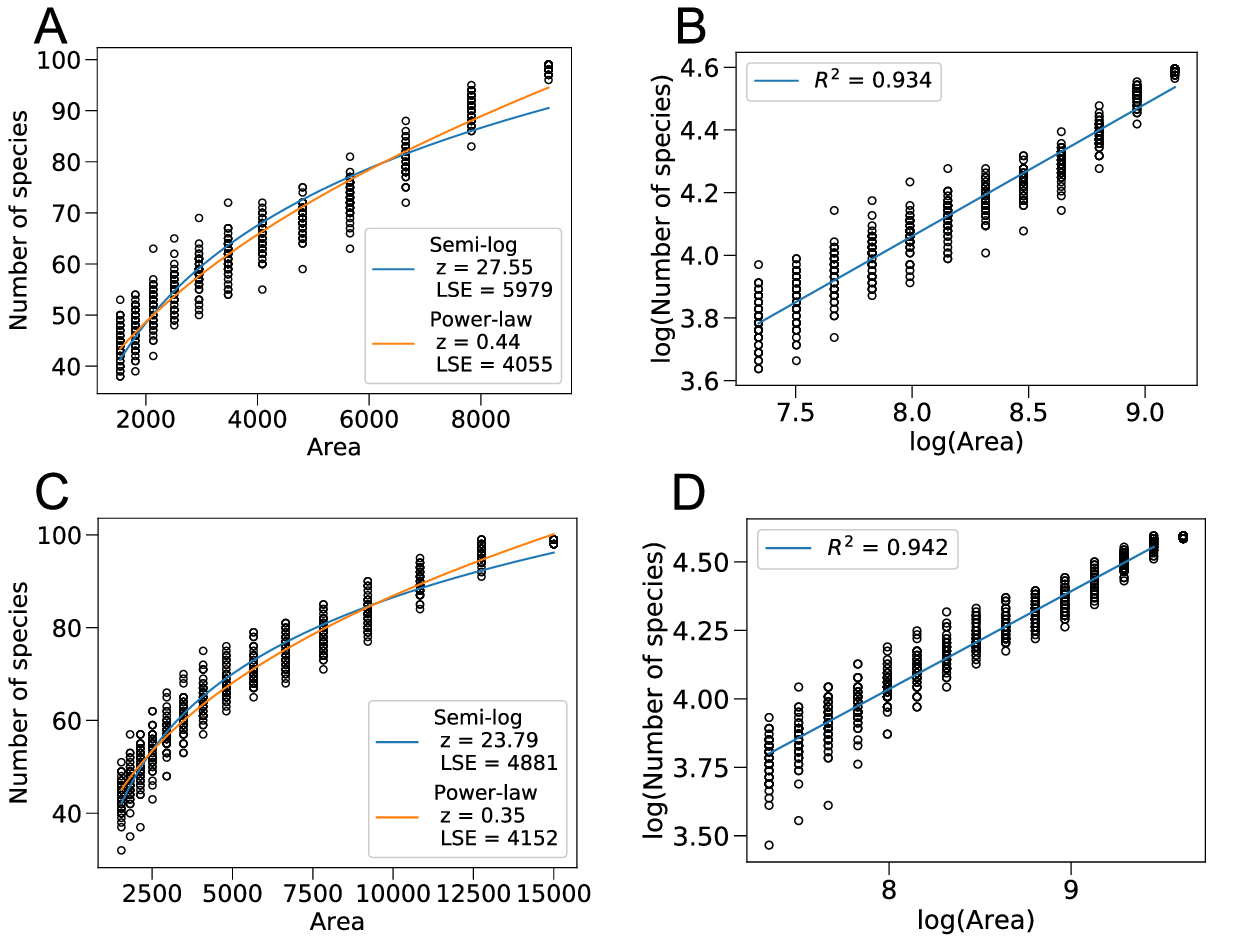}
    \caption{ Species area plots demonstrating the better fit of power law SAR for intermediate values of immigration rates. Panels A and C show the fits for $\lambda=0.1$ and $\lambda=0.01$ respectively for 50 instances of the interaction matrix. Panels B and D correspond to the respective linear regressions on log-log plots. The interaction strength mean and variance are -1 and 0.2 respectively. $A_{init}$ = 15000.}
    \label{fig3}
\end{figure}

Our model -- in its simplest form -- supports the semi-log relationship that is also widely reported in literature (\cite{drakare2006imprint}). Our analysis suggests that varying levels of immigration lead to different functional forms of the SAR. We start with a very low value of $\lambda$ and progressively increase it to check the resulting SAR. For very low immigration rates, the semi-log relationship is supported (see Fig. S1 in Supplementary Material) as seen in the scenario without immigration (Fig. 2). However, there exists an intermediate regime best characterised by a power law (Fig. 3). This form of the SAR also lacks the upper asymptote that we observed in the semi-log fit (Fig. 2). Interestingly, using area in the immigration term instead of its square root does not change the above results (Fig. S2 in Supplementary Material).

 \begin{figure}[!htb]
    \centering
    	\includegraphics[width =\linewidth]{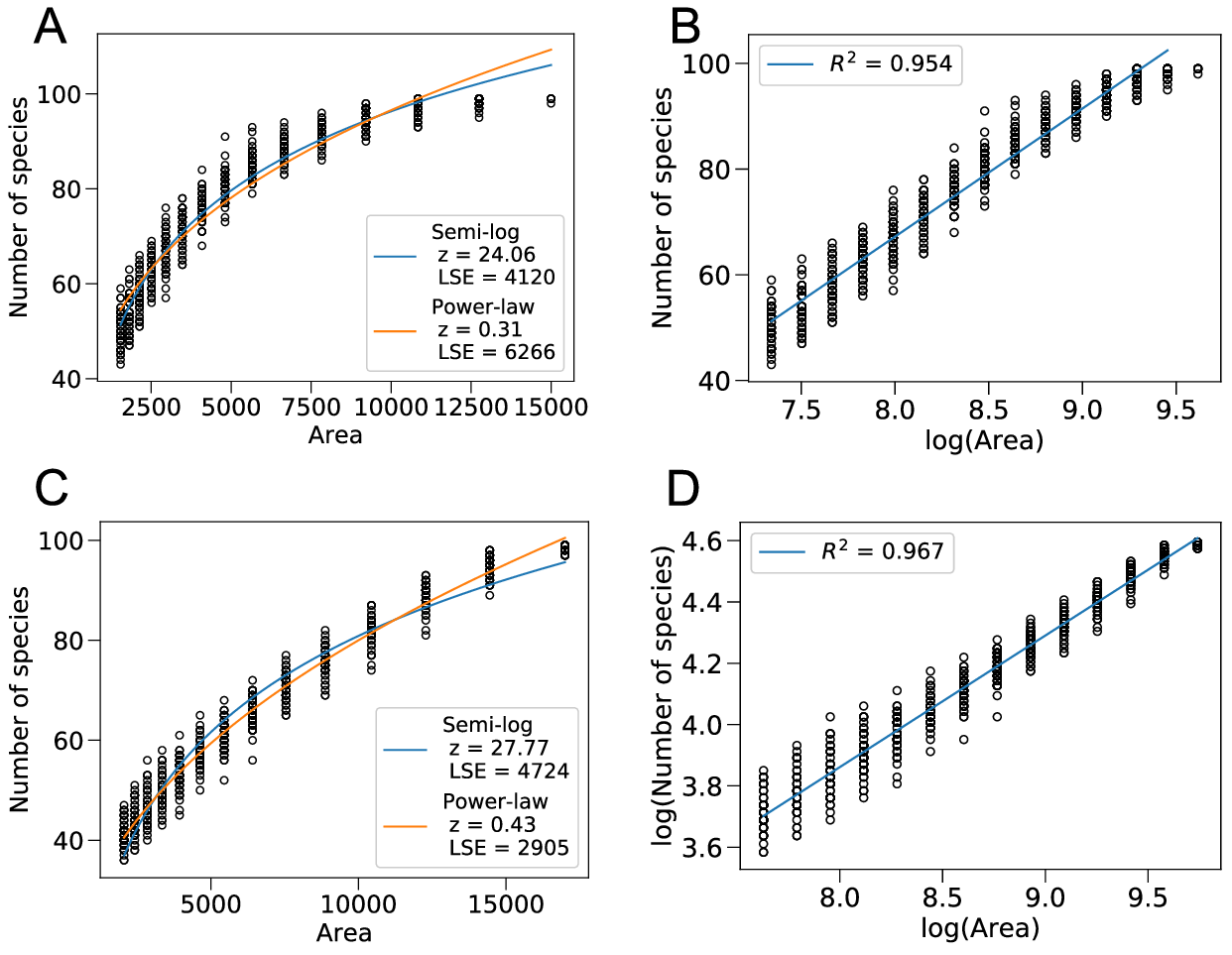}
    \caption{ SAR plots for exponentially distributed interactions with two different rate parameters. All plots correspond to $\lambda$ = 0.01 and connectance = 0.1, where the entries of the interaction matrix are chosen randomly as an Erdős-Rényi graph. The semi-log form is better supported for rate parameter =  0.5, as demonstrated by the estimates in A ($A_{init}$ = 15000). Plots C shows the fits for rate parameter = $0.25$, where the power law performs better ($A_{init}$ = 20000). (B) Linear regression on a semi-log plot using the same simulated data as in panel A. (D) Log-log plot showing the corresponding linear regression for data in C. }  
    \label{fig4}
\end{figure}

The level of skew towards weak interactions strongly influences SAR shape. Given the same immigration level, a higher skew towards weak interactions favours a semi-log relationship (Fig. 4, Fig. S3 in Supplementary Material). This result does not change for fat-tailed distributions such as the Pareto distribution in the regime where stable solutions exist (Fig. S3 in Supplementary Material). When studied for communities with low connectances, the SARs thus obtained have lower slopes as in most natural communities (Fig. 4).

\section{Discussion}

Much attention has been devoted to explain the power law form of the SAR. Our results show the emergence of two most widely observed forms of SAR through differences in immigration rates and skewness towards weak interactions (Fig. 5). We use a simple spatially-implicit model that demonstrates properties of SARs in island-like systems where inter-island immigration is very low. Our analysis supports the view that community dynamics could result in the emergence of such spatial patterns. In general, since we relate area to the interaction strength parameter, the SARs seem to stem from the scaling of interactions for species within the same trophic level. 

\begin{figure}[!htb]
   \centering
       \includegraphics[width =\linewidth]{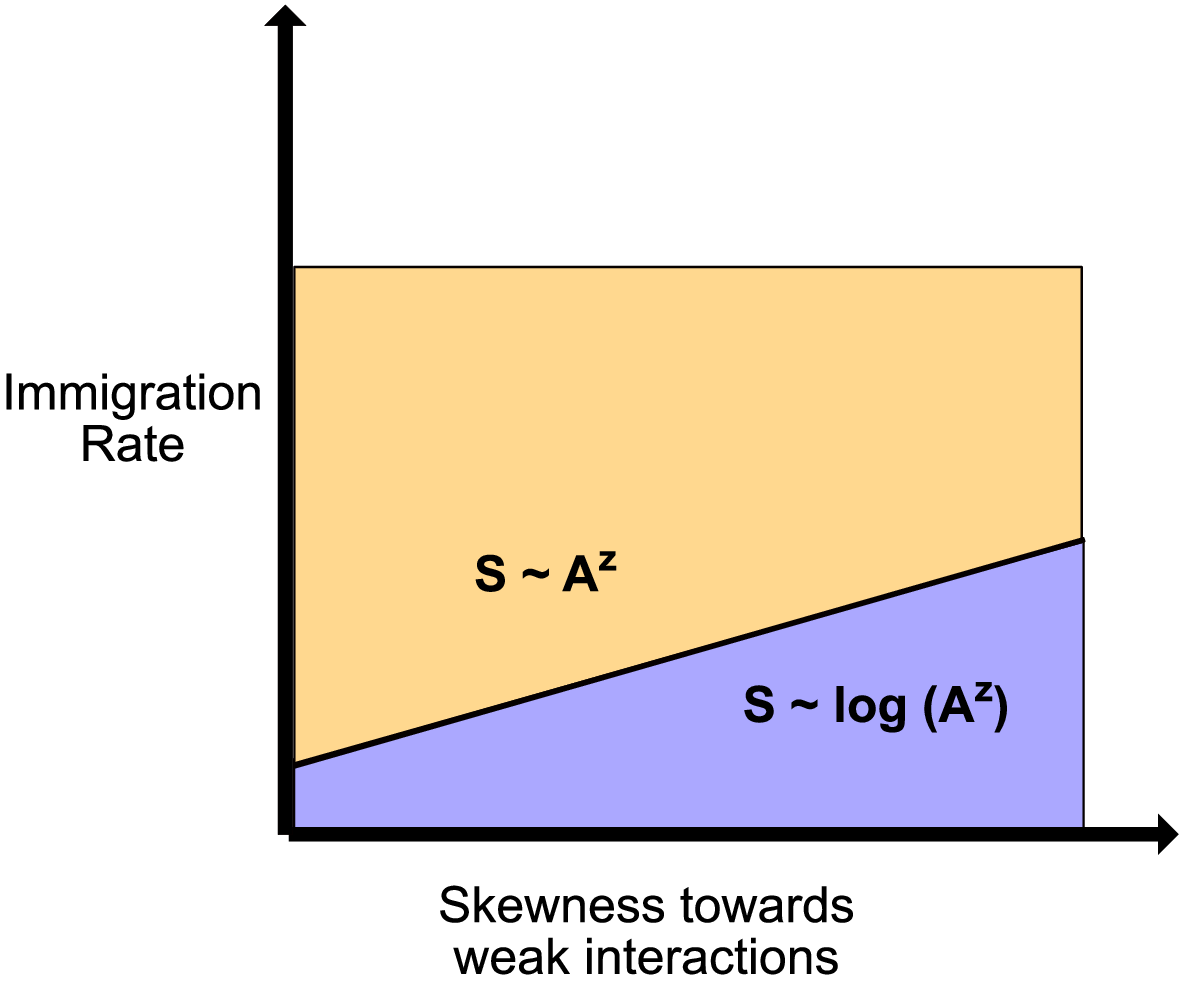}
      \caption{Immigration rates and skewness towards weak interactions determine SAR forms. Semi-log relationship dominates in the absence of immigration. Higher immigration rates from a source pool result in power law relationships but these could shift to semi-log SARs if the relative proportion of weak interactions is increased. $S$, $A$ and $z$ represent the number of species, area and the scaling law exponent respectively. }           
      \label{fig5}
\end{figure}

In addition to immigration rates and skewness towards weak interactions, connectance also influences SAR slopes. If all other parameters are kept the same, then communities with lower connectance result in lower SAR slopes (Fig. 4, Fig. S3 in Supplementary Information). We also find that higher immigration rates correspond to higher SAR slopes, such that the number of surviving species fall off much more sharply with area for large areas. The SAR slopes we obtain are much more reasonable for choices based on realistic interaction networks (Fig. 4). We expect that some network structures could result in even lower slopes, but without much effect on the SAR form.

\subsection{ Immigration shapes SARs: The case of remote archipelagoes}
\label{sec_eg}

Our results have important implications for island systems, which we illustrate using two extensive empirical studies from the Southwest Pacific (\cite{diamond1972biogeographic, diamond1976species}). The Solomon archipelago in (\cite{diamond1976species}) is more than 600 km away from the 'source island' of New Guinea. The authors assume that intra-archipelago immigration rates are much higher than the immigration rates from the `source' island of New Guinea.

\begin{figure}[!htb]
    \centering
    	\includegraphics[width =0.6\linewidth]{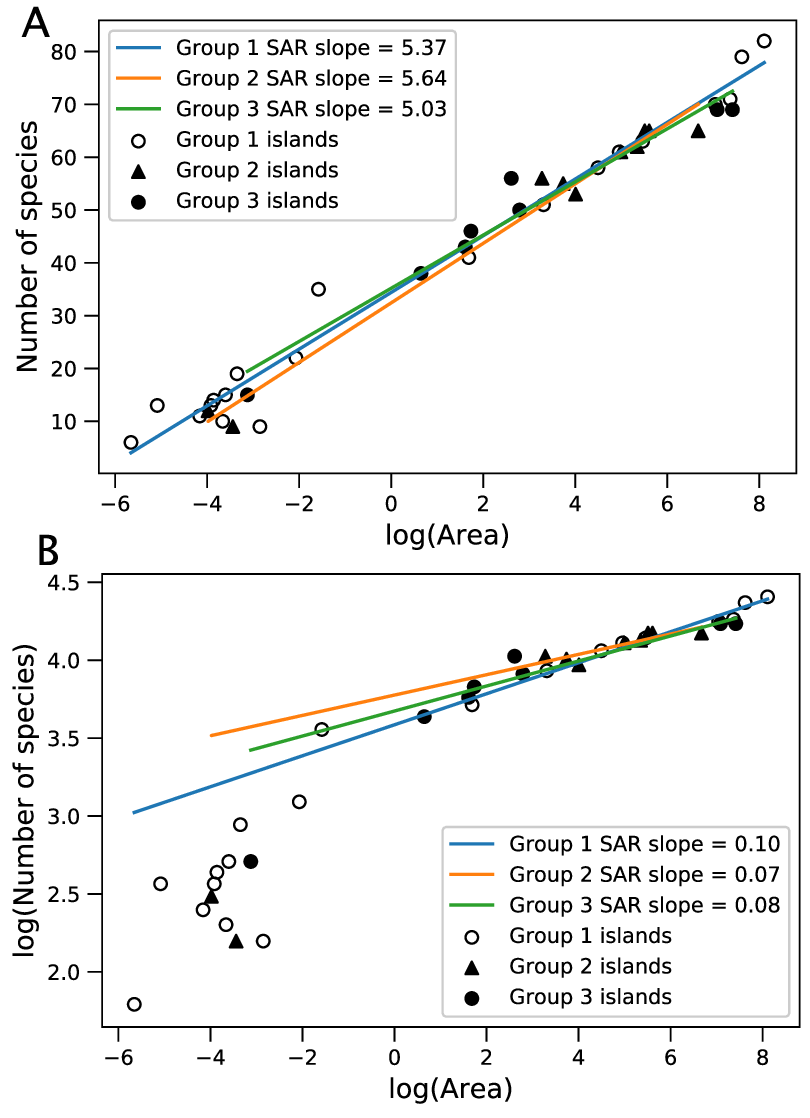}
    \caption{ SAR plots for three groups of non-isolated islands within the Solomon Archipelago. These groups differ in how the islands within them were connected during the Pleistocene period. The islands in Group 3 did not have any history of connections. The semi-log relationship shows a good fit to data (A). The $R^2$ values for the regression lines are 0.978, 0.982 and 0.955 for Group 1, 2 and 3 respectively. The slopes for the different groups are very similar. Panel B shows a clear departure from a power-law relationship for smaller areas. The linear regression lines indicate a good fit for islands larger than one square mile. In particular, the $R^2$ value for such islands in group 1 is 0.976 from the power-law SAR. This demonstrates that a naive inference  could support a power law, in spite of the islands spanning over four orders of magnitude in area ( $>$ 1 square mile).}  
    \label{fig6}
\end{figure}

They further plot the SAR for three groups of islands within the Solomon Archipelago, which supports a semi-log form. The slope of the SAR is nearly the same across these three groups of islands (Fig. 6). We surmise that the immigration of birds into an island is also balanced by emigration to other islands within the same group. In other words, the system is at a steady-state of zero or very low immigration within the archipelago. Thus, any effective immigration should emanate from the source island or from islands in other distant archipelagos. The large distance from these other sources implies that the net immigration rates to the Solomon islands are very low. In fact, the authors state that with increasing isolation of an archipelago, the SAR may shift in form from a power function to an exponential (semi-log). The species richness data from the Azores (\cite{whittaker2014functional}) as well as Andaman Islands (\cite{gooriah2020species}) also concurs with this claim (See Supplementary Appendix S2). The Azores archipelago includes only nine islands but has been extensively studied over the past many decades. Both relationship forms show good fits to the data primarily because of the small number of islands but the semi-log form is more predictive for smaller islands. All of these studies are consistent with our theoretical finding that low immigration rates lead to semi-log SARs. 

The dataset in (\cite{diamond1976species}) also has many islands smaller than a few square kilometres, which are usually absent in many SAR studies (\cite{lomolino2001towards}). Both forms of the SAR could show a very good (and similar) fit to data for larger island sizes (Fig. 6). As the authors point out, really small islands should be included in SAR analyses to conclusively identify the correct form of the relationship (\cite{diamond1976species}). 

Another study from islands that lie 5 to 300 miles from New Guinea, found a power-law SAR (\cite{diamond1972biogeographic}). Considering that these islands lie closer to the `source' island of New Guinea, the immigration rates are likely to be higher than those for the Solomon Archipelago. This lends support to our theoretical results on the incidence of power-law SARs for higher immigration levels. 

\section{Conclusion}

Given a random configuration of competitive species, we recover many known features of SARs while also identifying factors that might best explain the variation in these relationships. The two SAR forms might show similar fits to data for a large span of areas but their differences could be stark for smaller islands especially when immigration rates from a source pool are low. Our results imply semi-log relationships for low immigration rates, which are possible through factors such as remoteness of an archipelago as in (\cite{diamond1976species}). Assuming a power law SAR in such situations could mislead extinction scenarios since these would overestimate the species richness for smaller areas. It is extremely important to investigate the effects of habitat loss, especially on small islands in distant archipelagoes, given that islands have witnessed disproportionately large number of extinctions (\cite{loehle2012historical, spatz2017globally}). We hope that our study prompts empirical studies to systematically evaluate the effects of immigration and community structure on species-area relationships.

\begin{acknowledgements}
We would like to thank Susanne Pettersson for useful discussions. AV is especially grateful to Akshay Surendra for detailed feedback on the manuscript.
\end{acknowledgements}

\bibliographystyle{spbasic}      

\bibliography{Spatial_GLV}{}   

\renewcommand{\theequation}{S\arabic{equation}}
\renewcommand{\thetable}{S\arabic{table}}
\renewcommand{\thesection}{S\arabic{section}}
\renewcommand{\thefigure}{S\arabic{figure}}
\setcounter{equation}{0}  
\setcounter{figure}{0}
\setcounter{table}{0}
\setcounter{section}{0}

\newpage{}

\section{Additional Methods}

To set up a system describing the dynamics of many species through our model, we choose normally distributed per capita growth rates, carrying capacities, and entries of the interaction matrix (unless explicitly specified). The growth rates are distributed around a positive mean such that there is a very low probability of having species with negative growth rates. Similarly, the random interactions are distributed around a negative mean such that almost all interactions are negative. Choosing carrying capacities from uniform or lognormal distributions is also consistent with the results that we report in this paper. 

The model is numerically solved for different areas for which stable equilibria exist. For a given area, this translates to numerically simulating the GLV system for many time steps until it converges to the stable fixed point of equation 2 in main text. With regards to the parameter choices, we tried to maximize the possible area values for which each of the numerical experiments could be run while ensuring that all solutions converge to stable equilibria. With regards to immigration, a species is considered extinct if its abundance falls below $10^{-5}$ for a given area. Initial abundances are drawn from a normal distribution (mean = 10, standard deviation = 0.1) but the final community does not depend on these abundances especially when stable equilibria exist. We simulate over successively smaller areas such that consecutive areas are related by a factor of 0.85. The simulations in figure 2 are an exception (we fix this factor as 0.75), partly because stable solutions exist for more number of iterations here. 

\newpage{}

\section{SAR for bird data from the Andaman and Azores islands}

Andaman Islands is an isolated archipelago comprising of many islands in the Bay of Bengal. Here we show that the bird dataset from this archipelago (\cite{gooriah2020data}) shows a better fit to the semilog SAR (Fig. S5), which is particularly evident at smaller areas. 

The Azores archipelago is a well-studied region that is considerably distant from the mainland but it comprises of only nine islands. We used the spider and beetles dataset reported in \cite{whittaker2014functional}. The parameter estimation from the linear regression and the non-linear least squares differs for the power law -- in part because of the low number of islands. Therefore, we also perform a model averaging using the R package `sars' (\cite{matthews2019sars}). The corrected Akaike Information Criterion (AICc) weights support the semi-log form, in agreement with the non-linear least squares method. These results also hold when only the indigenous (non-exotic) species are considered.

\newpage{}

\section{Supplementary Figures}

\begin{figure}[h!]
        \includegraphics[width =\linewidth]{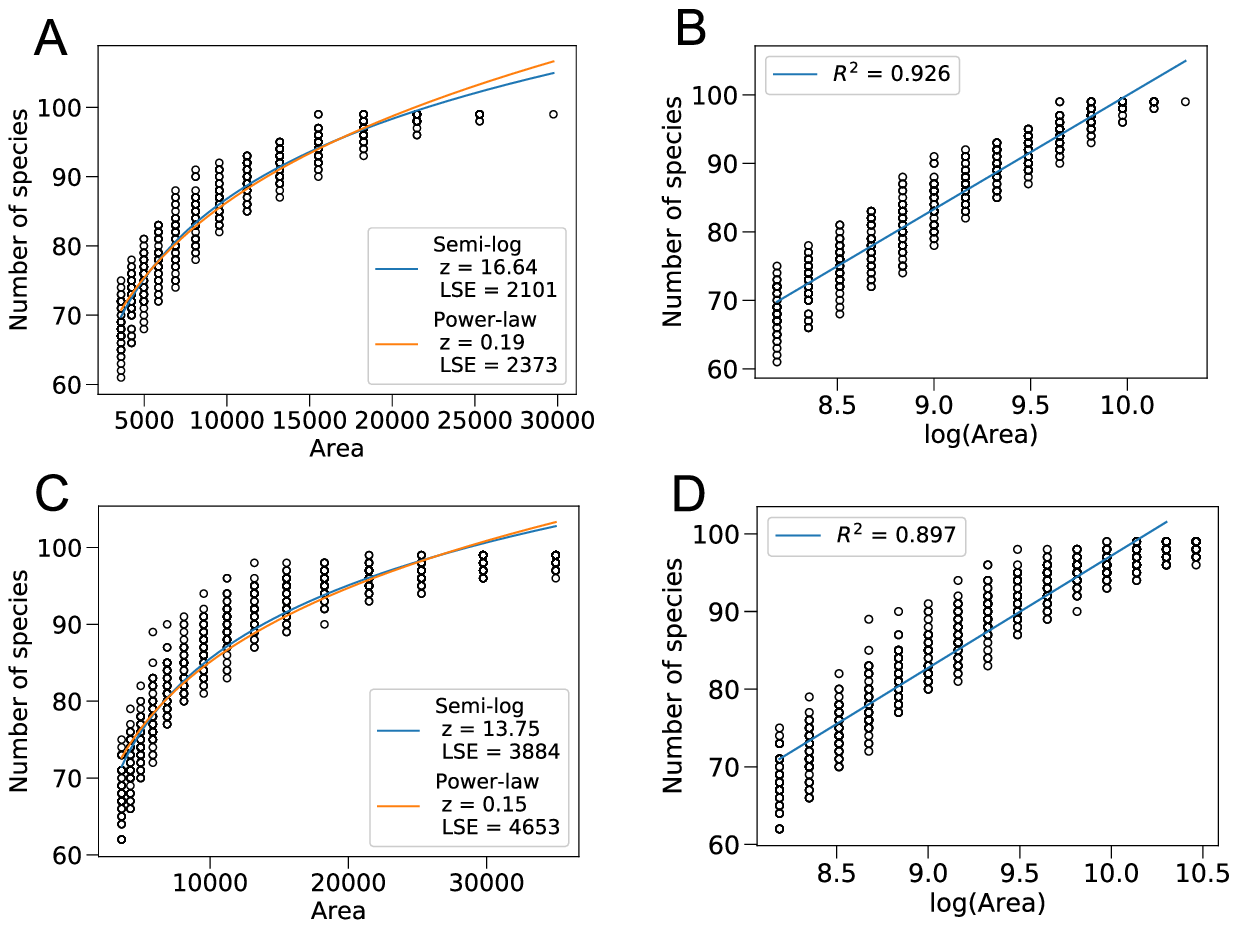}
        \caption{The semi-log form supports the data better for low levels of immigration (A and C). $\lambda=0.001$ for A and $\lambda=0.0001$ for the C. Panels B and D show the linear regressions on semi-log plots for data from A and C respectively. The simulations were generated using $A_{init}$ = 35000 and interaction matrix with mean=-1 and variance=0.2.}
	\label{figs1}
\end{figure}

\begin{figure}[h!]
        \includegraphics[width =\linewidth]{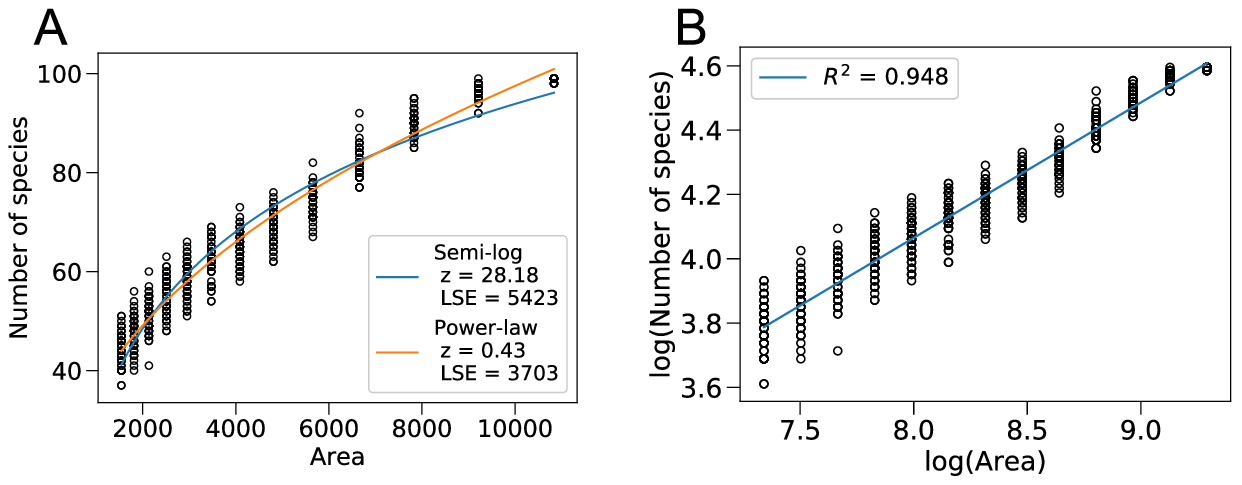}
	\caption{ SAR plots when immigration term is $  \lambda e^{-\beta /A } $ where $\beta = 30000$ and $\lambda = 5 \times 10^{-5}$. We obtain the power-law SAR for the specified $\lambda$ when the immigration decreases exponentially with area, as shown in A. The semi-log relationship dominates if $\lambda$ is any lower. (B) Log-log plot with the linear regression using the same data as in A. $A_{init}$ = 15000}
	\label{figs2}
\end{figure}

\begin{figure}[h!]
	\includegraphics[width =\linewidth]{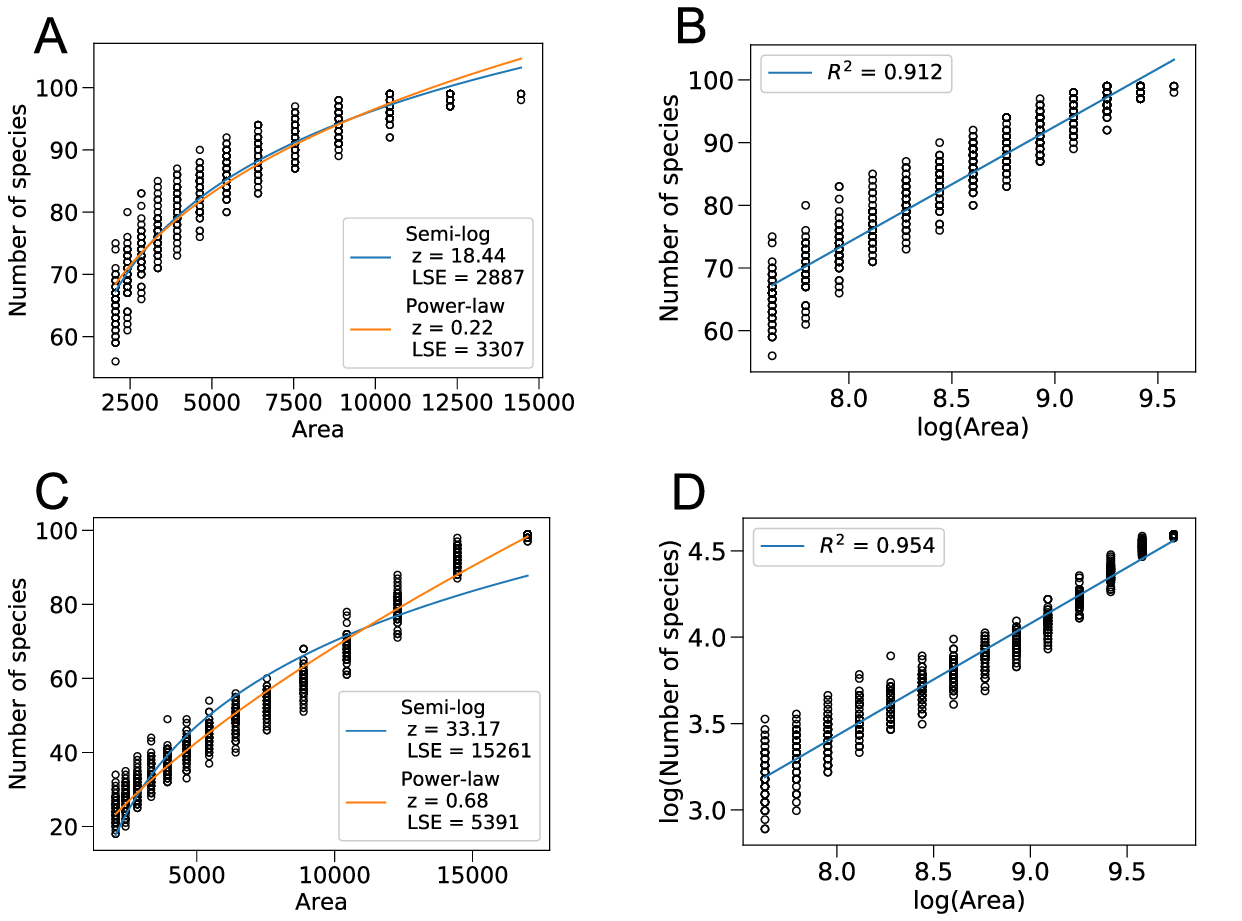}
	\caption{ SAR plots for $\lambda = 0.01$, connectance = 1 where the interactions are drawn from an exponential distribution. Panel A corresponds to rate parameter = 2 that results in a better fit for the semi-log form. Panel C implies a better fit for the power law where rate parameter = 0.5. Panels B and D show the linear regressions corresponding to data in A and C respectively. $A_{init}$ = 20000 for all cases here.}
	\label{figs3}
\end{figure}

\begin{figure}[h!]
        \includegraphics[width =\linewidth]{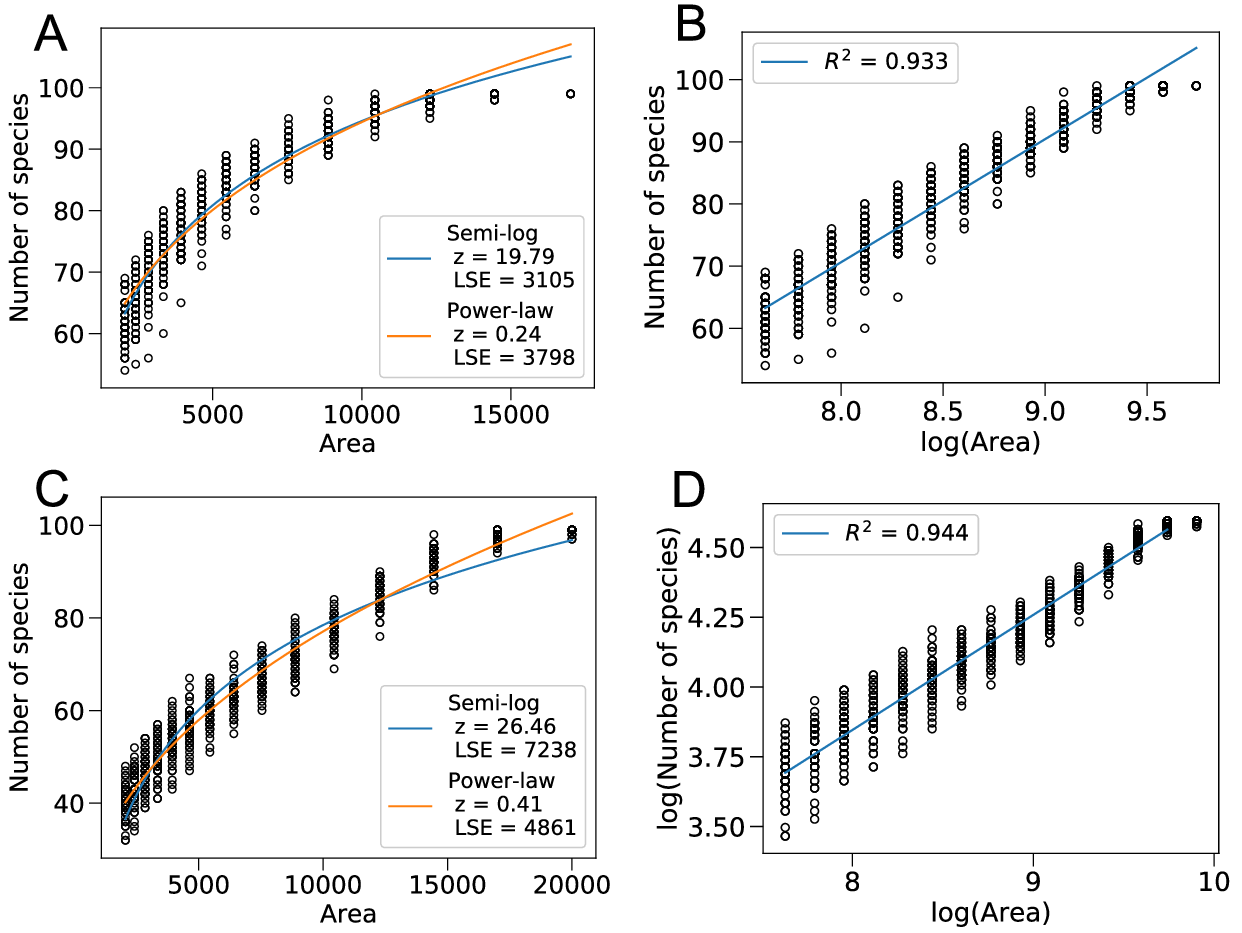}
	\caption{ SAR plots for $\lambda = 0.01$, connectance = 1 where the interactions are drawn from a Pareto distribution. Panels A and B correspond to shape parameter = 3. A semi-log function fits the simulated data in A better --  panel B shows the corresponding linear regression. The least-squares error in panel C implies a better fit for the power law where shape parameter = 2. Panel D is a log-log plot showing the corresponding linear regression. $A_{init}$ = 20000 for all cases here.}
	\label{figs4}
\end{figure}

\begin{figure}[h!]
        \includegraphics[width =\linewidth]{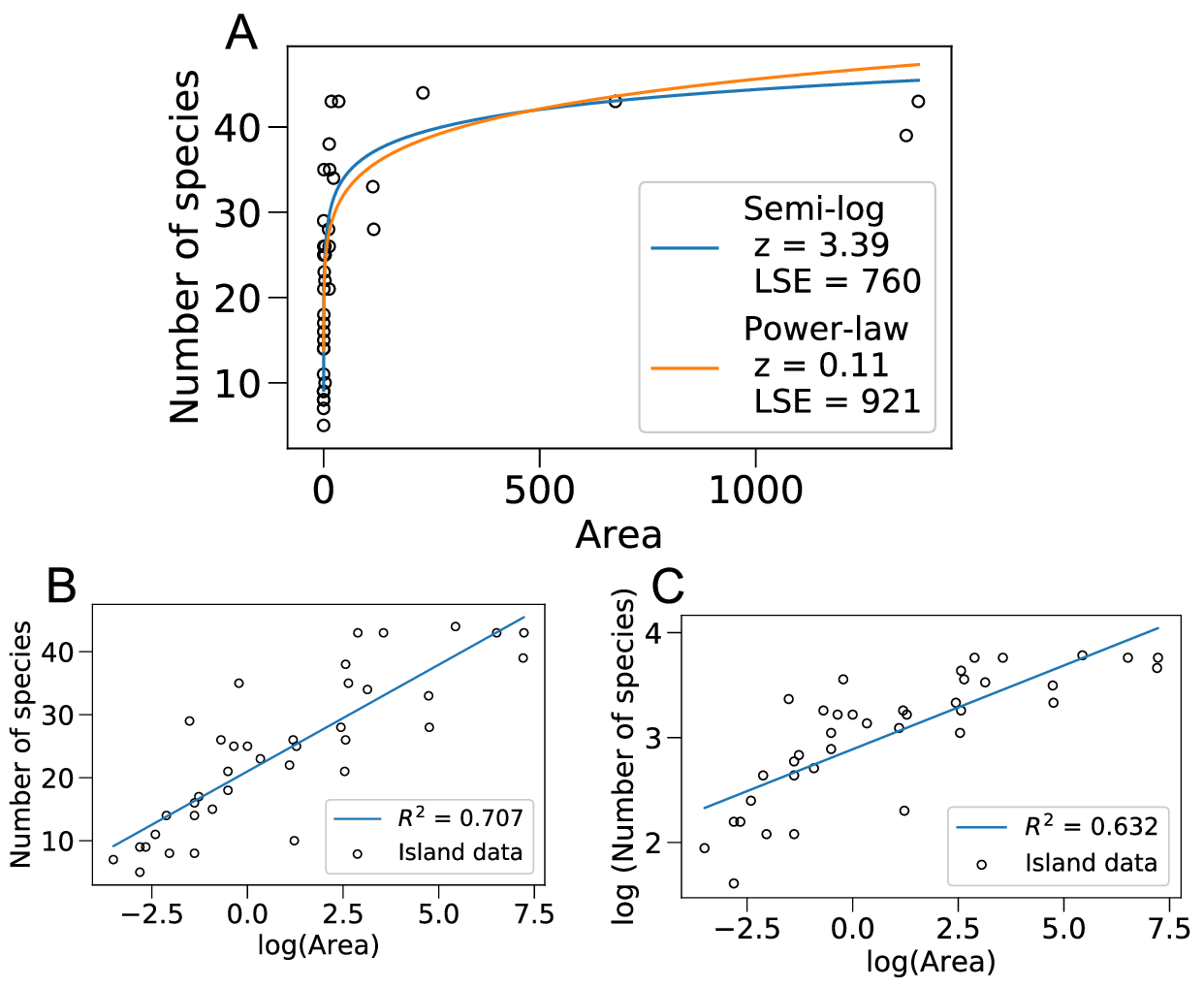}
	\caption{ (A) The semi-log form shows a better fit to the bird data from Andaman Islands, shown using non-linear least squares errors. (B) and (C) Plots showing linear regression fits on a log-log and semi-log plot respectively. Notice the larger deviation at lower areas in the log-log plot. }
	\label{figs5}
\end{figure}

\begin{figure}[h!]
        \includegraphics[width =\linewidth]{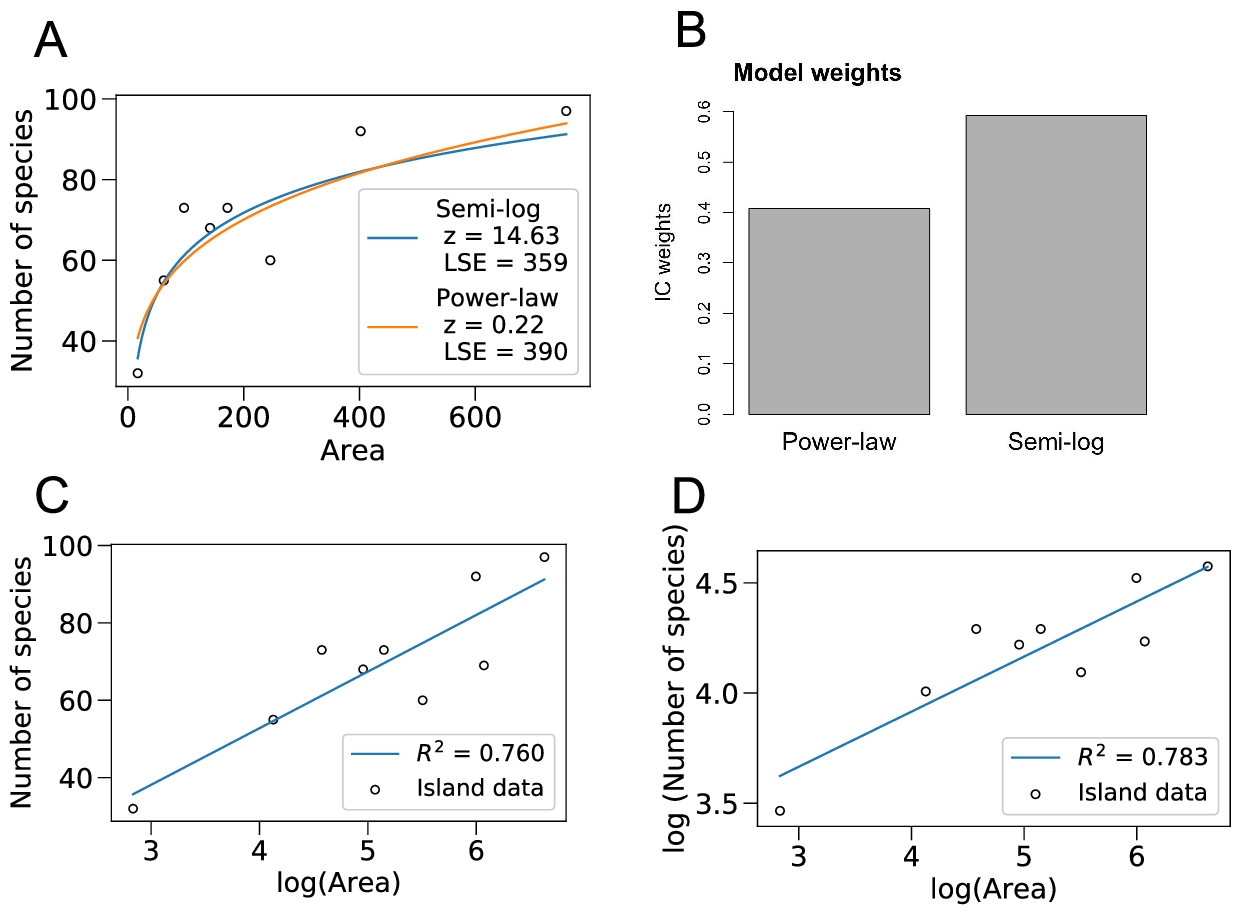}
	\caption{ Spider diversity data from Azores Islands. (B) We perform model averaging over the two functional forms to compare the model fits using AICc weights. The semi-log form performs slightly better but the differences are hard to compare using other methods (A, C and D) . }
	\label{figs6}
\end{figure}

\begin{figure}[h!]
        \includegraphics[width =\linewidth]{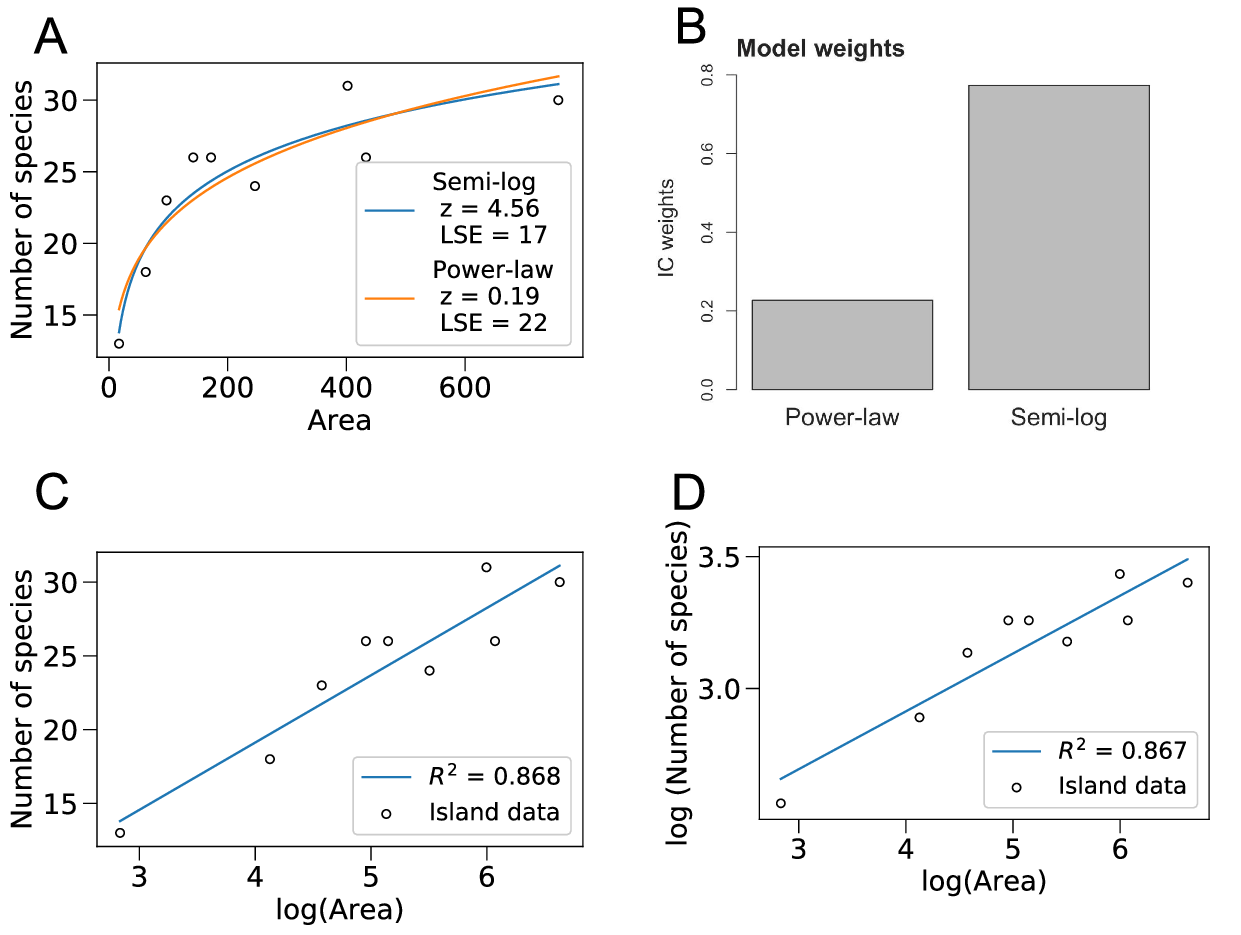}
	\caption{ Indigenous spider diversity data from Azores Islands. (B) We perform model averaging over the two functional forms to compare the model fits using AICc weights. The semi-log form performs slightly better but the differences are not hard to assess using other methods (A, C and D). }
	\label{figs7}
\end{figure}

\begin{figure}[h!]
        \includegraphics[width =\linewidth]{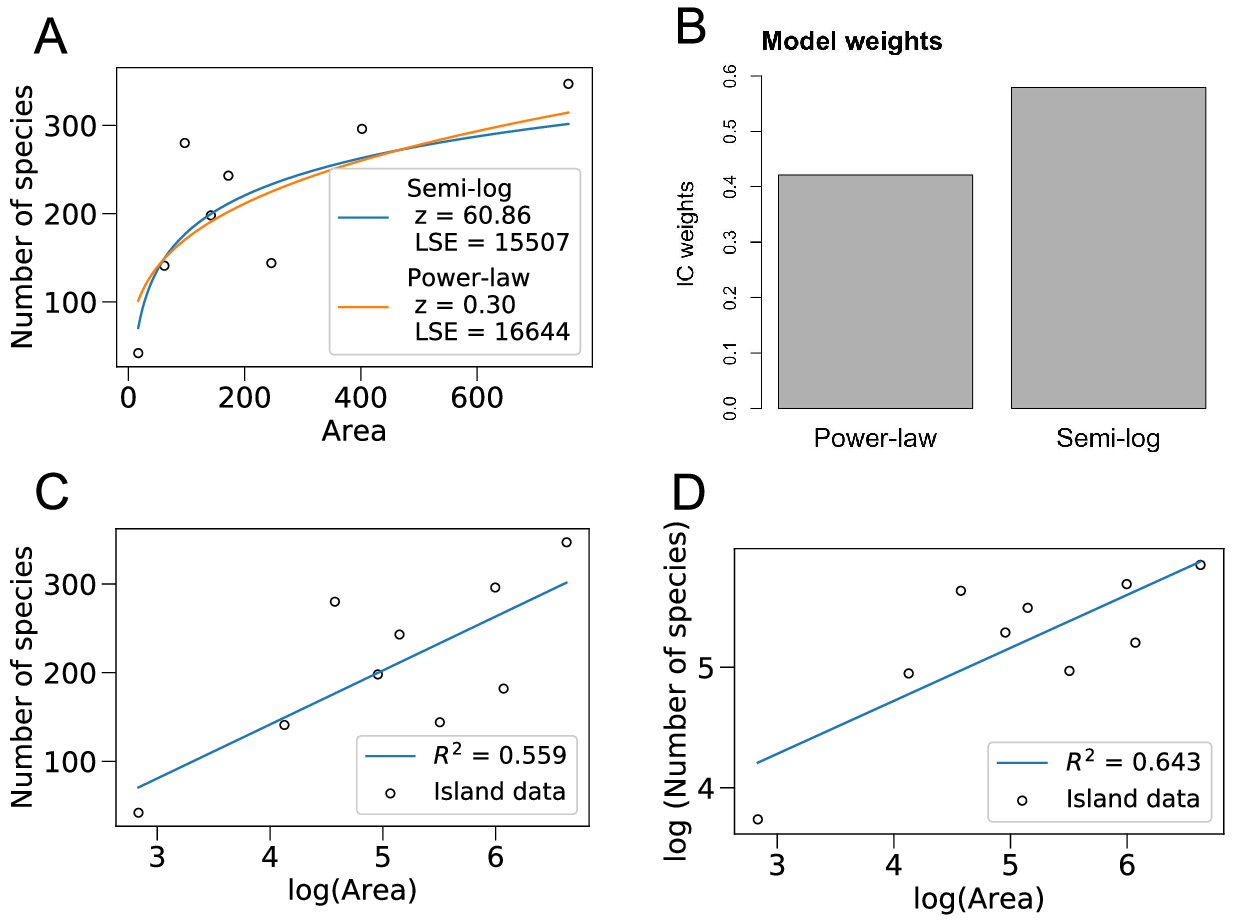}
	\caption{ Beetle diversity data from Azores Islands. (B) We perform model averaging over the two functional forms to compare the model fits using AICc weights. The semi-log form performs slightly better but the differences are hard to compare using other methods, especially because the differences in parameter estimates for the power-law fit are considerable. (A, C and D) . }
	\label{figs8}
\end{figure}

\begin{figure}[h!]
        \includegraphics[width =\linewidth]{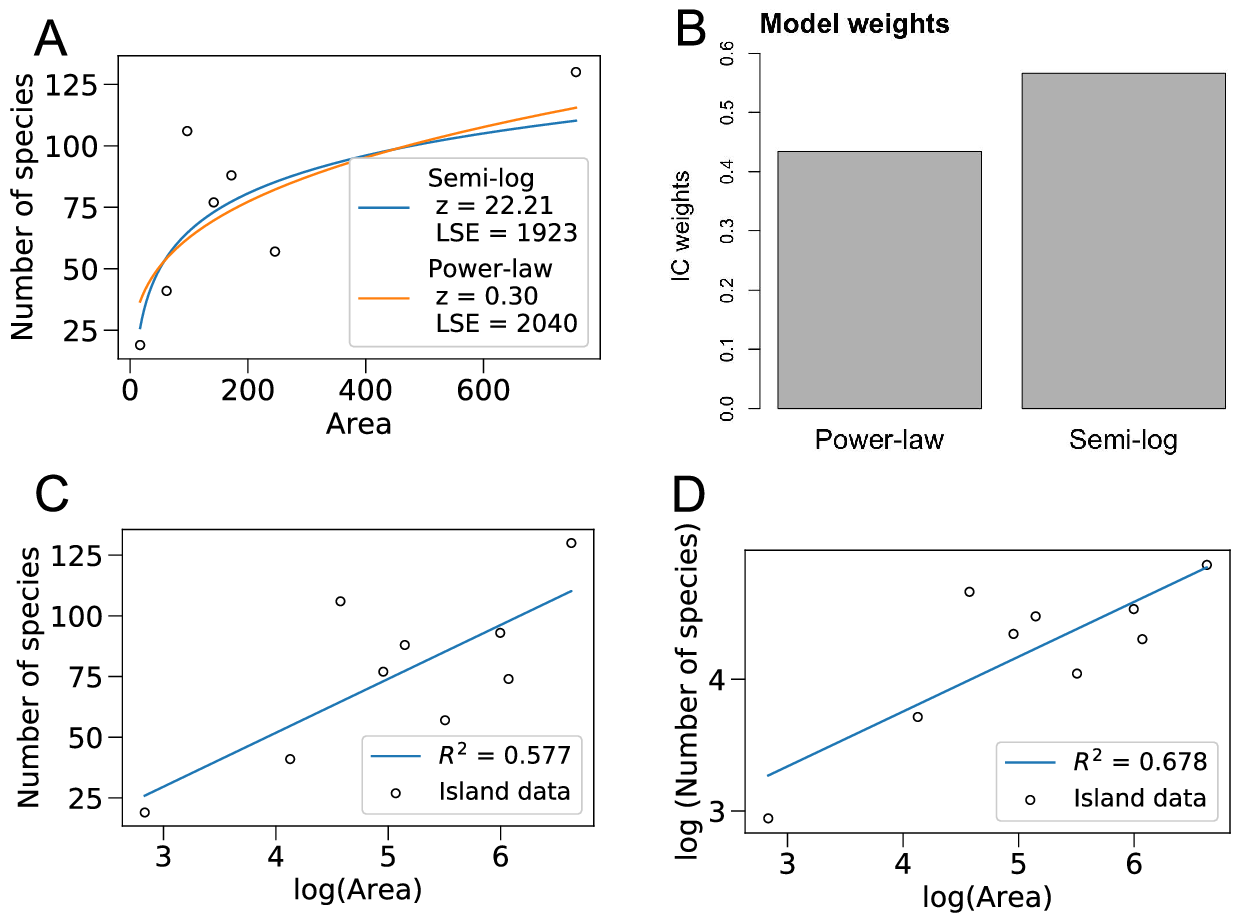}
	\caption{  Diversity data for indigenous beetles from Azores Islands. (B) We perform model averaging over the two functional forms to compare the model fits using AICc weights. The semi-log form performs slightly better but the differences are not hard to assess using other methods, especially because the differences in parameter estimates for the power-law fit are considerable. (A, C and D). }
	\label{figs9}
\end{figure}

\end{document}